\documentclass[aps,twocolumn,prb,showpacs]{revtex4-1}
\usepackage{amsfonts}
\usepackage{amsmath}
\usepackage{amssymb}
\usepackage[final]{graphicx}
\begin{document}
\title{Anomalous density of states in multiband superconductors near Lifshitz transition}

\author{A. E. Koshelev }
\affiliation{Materials Science Division, Argonne National Laboratory, Argonne,
Illinois 60439}
\author{K. A. Matveev}
\affiliation{Materials Science Division, Argonne National Laboratory, Argonne,
Illinois 60439}
\date{\today }
\begin{abstract}
  We consider a multiband metal with deep primary bands and a shallow
  secondary one.  In the normal state the system undergoes Lifshitz
  transition when the bottom of the shallow band crosses the Fermi
  level. In the superconducting state Cooper pairing in the shallow
  band is induced by the deep ones. As a result, the density of
  electrons in the shallow band remains finite even when the bottom of
  the band is above the Fermi level. We study the density of states in
  the system and find qualitatively different behaviors on the two
  sides of the Lifshitz transition. On one side of the transition the
  density of states diverges at the energy equal to the induced gap,
  whereas on the other side it vanishes. We argue that this physical
  picture describes the recently measured gap structure in shallow
  bands of iron pnictides and selenides.
\end{abstract}
 \maketitle

Recent discovery of superconductivity in iron-based materials is one of the most
important developments in modern condensed matter physics
\cite{KamiharaJACS08,PaglioneNPhys10,JohnstonAdvPhys10}. In addition to high
transition temperatures, these materials have several exciting features including
the interplay of superconductivity with spin-density wave order, a possibility of
electronic mechanism of pairing, and the formation of unconventional superconducting
state \cite{HirschfeldRPP11,ChubukovAnnual12}.  The new physics is observed in a
wide variety of materials, whose properties can be fine-tuned by doping.

A common feature of iron-based superconductors is the multiple-band electronic
structure. Some of these bands are very shallow, with Fermi energies of several
millielectronvolts, and may be depleted with doping or pressure
\cite{LiuNatPhys10,LiuPhysRevB11,miao_2014}. Such a qualitative change of the
Fermi-surface topology is known as the Lifshitz transition \cite{Lifshitz,
BlanterPhysRep94}. Superconducting properties of shallow bands have a number of
interesting features. For example, if was recently demonstrated by ARPES technique
that in the compounds FeSe$_{1-x}$Te$_{x}$ \cite{LubashevskyNat2012,OkazakiSciRep14}
and LiFe$_{1-x}$Co$_{x}$As \cite{miao_2014} the minimum gap for the shallow band is
realized at zero momentum, rather than at the Fermi surface, as expected in the
standard BCS theory. Furthermore, in Refs.~\cite{OkazakiSciRep14,miao_2014} the
superconducting gap was observed on the side of the Lifshitz transition where the
band would have been empty in the normal state. These observations were interpreted
as a manifestation of the Bose-Einstein condensation of electron pairs formed as a
result of strong electron-electron attraction \cite{GiorginiRevModPhys08}.

The goal of this paper is to present an alternative physical scenario based on the
notion that the superconductivity in the shallow band may be induced by deep bands
via pair-hopping. In the case when superconducting pairing is dominated by the deep
bands, the gap parameter in the shallow band is primarily determined by the
properties of deep bands and may be understood in the mean-field approximation.
Within our scenario the superconducting state in the shallow band is not a result of
the Bose-Einstein condensation even though the gap may be larger than the Fermi
energy. The influence of the shallow band on the transition temperature and other
global properties is typically weak due to its small density of states
\cite{MakarovBaryakhtarZheTF65}. However, its superconducting properties are very
different from the conventional BCS state due to strong violation of the
particle-hole symmetry.

It is interesting to note that superconductivity changes the nature of the Lifshitz
transition. In particular, the carrier density in the shallow band remains nonzero
on both sides of the transition. Finite density appears because the particle-hole
mixing generates a finite density of states (DoS) in the energy range where
normal-state DoS was zero leading to appearance of a long tail in
superconducting-state DoS. The only qualitative change at the transition is
modification of the excitation spectrum. At the critical value of the chemical
potential the minimum energy of excitations moves to the band center, as observed
experimentally \cite{LubashevskyNat2012}. This change is reflected in the shape of
DoS which changes dramatically as a function of the chemical potential. While on one
side of the transition DoS diverges at the gap energy as predicted by the BCS
theory, on the other side it vanishes at the gap energy.

\begin{figure}[t]
\vspace{-0.03in}
\begin{center}
\includegraphics[width=3.0in]{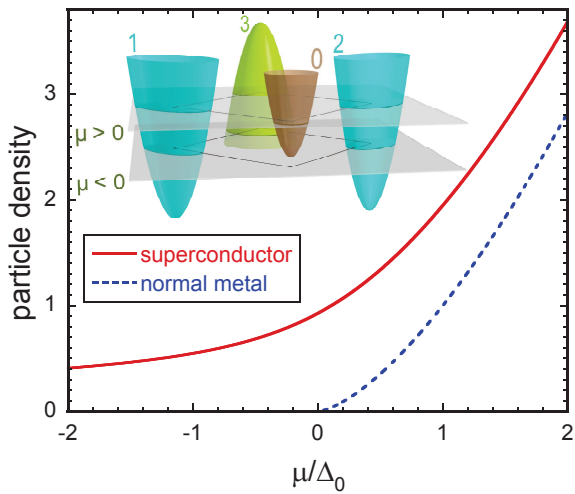}
\end{center}
\vspace{-0.2in}
\caption{
The dependence of the particle density in shallow band on the chemical potential for
normal metal and superconductor. The density unit is $n(\Delta_{0})$. While in the
normal state the density vanishes at the Lifshitz transition point, $\mu=0$, in the
superconducting state it remains finite for all $\mu$. The inset illustrates
electronic spectrum of a multiband metal with the shallow band.} \label{Fig-n-mu}
\vspace{-0.12in}
\end{figure}
We consider a superconductor with $M$ deep bands and one shallow band, as
illustrated by the inset in Fig.\ \ref{Fig-n-mu}. The starting point for our
discussion of the superconductivity in the shallow band is the BCS Hamiltonian
\begin{equation}\hspace{-0.027in}
H_{0}\!=\!\sum_{\mathbf{p},\sigma}\xi_{\mathbf{p}}
a_{\mathbf{p},\sigma}^{\dagger}a_{\mathbf{p},\sigma}
\!+\!\sum_{\mathbf{p}}
\Delta_{0}\left(  a_{\mathbf{p},\uparrow}^{\dagger}a_{-\mathbf{p},\downarrow
}^{\dagger}\!+\!a_{-\mathbf{p},\downarrow}a_{\mathbf{p},\uparrow}\right)\!,\! \! \!
\label{hamilt}
\end{equation}
Here the operator $a_{\mathbf{p},\sigma}$ destroys an electron in the shallow band
with momentum $\mathbf{p}$ and spin $\sigma$, and
\begin{equation}
  \label{eq:2}
  \xi_{\mathbf{p}}=p^{2}/(2m_0)-\mu,
\end{equation}
where the chemical potential $\mu$ is measured from the bottom of the shallow band.
For definiteness we assumed an isotropic electronlike shallow band, i.e., $m_0>0$.
The point $\mu\!=\!0$ corresponds to the Lifshitz transition in the normal state at
which this band becomes depleted, see inset in Fig.\ \ref{Fig-n-mu}. The pairing
amplitude $\Delta_0$ is induced in the shallow band by Cooper pair exchange with the
deep bands. At zero temperature it is given by
\cite{MultibandBCS,MoskalenkoFMM59,GeilikmanFTT67}
\begin{equation}
\Delta_{0}=\sum_{j=1}^{M}V_{0,j}\nu_{j}\Delta_{j}\ln\frac{W}{\Delta_{j}}.
\label{ShallGapEq}
\end{equation}
Here $V_{0,j}$ are the amplitudes of pair hopping between the shallow and deep
bands, $\Delta_{j}$ and $\nu_{j}$ are the values of the gap and normal DoS in the
deep bands.  The value of the cut off parameter $W$ depends on the mechanism of
Cooper pairing; equation (\ref{ShallGapEq}) only assumes that $W\gg \Delta_{j}$.

We emphasize that the validity of the mean-field equation (\ref{ShallGapEq})
requires conditions that $\Delta_{j}$ are much smaller than the Fermi energies
$E_{F,j}$ for the deep bands, while the relation between $\Delta_0$ and
$E_{F,0}\!\equiv \! \mu$ may be arbitrary.  Note that the sum in the right hand side
of Eq.~(\ref{ShallGapEq}) excludes the term $j\!=\!0$ corresponding to the
contribution to pairing from the shallow band. In the limit $\mu \!\gg\! \Delta_0$
this term also has the mean-field form $V_{0,0}\nu_{0}\Delta_0\ln(\mu/\Delta_0)$.
However, since we are interested in the regime when the bottom of the shallow band
is close to the Fermi level, the density of states $\nu_0$ is small, and such
contribution is negligible compared to those of other bands \footnote{The scenario
we consider here is distinctly different from
  the situation when the shallow band dominates the Cooper pairing so
  that its depletion destroys superconducting state
  \cite{LiuPhysRevB11,CooleyPhysRevLett05,InnocentiPhysRevB10}.}. Even
though at $\mu\! \sim\! \Delta_0$ the shallow-band contribution can not be described
by the mean-field approach, it remains small. As a result, all the gap parameters
$\Delta_j$, including $\Delta_0$, can be assumed to be independent of $\mu$.

Diagonalization of the Hamiltonian (\ref{hamilt}) with the Bogoliubov transformation
gives the standard quasiparticle spectrum \vspace*{-0.1in}
\begin{equation}
E_{p}=\sqrt{\xi_{p}^{2}+\Delta_{0}^{2}}. \label{Spectrum}
\end{equation}
The electron and hole contributions to the Bogoliubov wave function of
quasiparticles are determined by the coherence factors
\[
u_{p}^{2} \!=\frac{1}{2}\left(  1+\frac{\xi_{p}}{E_{p}}\right) ,\ v_{p}^{2}
\!=\frac{1}{2}\left(  1-\frac{\xi_{p}}{E_{p}}\right)  .
\]
We emphasize that in our case these standard mean-field results are valid for any
relation between $\mu$ and $\Delta_{0}$ including the region of empty band in the
normal state, $\mu\!<\!0$.

The Lifshitz transition in the normal metal is characterized by non-analytic
behavior of the particle density as a function of the chemical potential
\cite{Lifshitz}. Indeed, the density of particles in the shallow band at zero
temperature vanishes at $\mu<0$,
\begin{equation}
  \label{eq:normal_density}
  n(\mu)=\frac{(2m_0\mu)^{3/2}}{3\pi^{2}}\,\theta(\mu).
\end{equation}
Here $\theta(x)$ is the unit step function.  To study the effect of
superconductivity on the Lifshitz transition we evaluate the particle density as
\[
n_{s}(\mu)=2\int\frac{d^3p}{(2\pi)^3}\,v_p^2.
\]
Introducing the natural scale $n(\Delta_{0})$ for the density, we present
$n_{s}(\mu)$ in the form
\begin{equation}
n_{s}(\mu)=n(\Delta_{0})\ G\left(  \mu/\Delta_{0}\right),  \label{ns}
\end{equation}
where the  function $G\left(  a\right)  $ is defined as
\[
G\left(  a\right)  =\frac{3}{4}\int\limits_{-\mathrm{arcsinh}a}^{\infty}
dx\exp(-x)\sqrt{a+\sinh x}.
\]
It can be expressed in terms of the full elliptic integrals $K(x)$ and $E(x)$ as
\begin{equation}
\label{eq:G}
G(a)  =\frac{1}{2}\left(  a^{2}+1\right)  ^{1/4}\left[
\frac{K\left[  r(a)\right]  }{\sqrt{a^{2}+1}+a}+2aE\left[  r(a)\right]
\right],
\end{equation}
where $r(a)=\frac12(\sqrt{a^{2}+1}+a)/\sqrt{a^{2}+1}$.

The dependences of particle densities on the chemical potential for normal and
superconducting states are shown in Fig.~\ref{Fig-n-mu}.  At $\mu\gg\Delta_0$ we use
the asymptotic behavior $G(a)\simeq a^{3/2}$ at $a\to\infty$ and find that
$n_{s}(\mu)$ approaches the normal-state density $n(\mu)$.  In the opposite limit
$-\mu\gg\Delta_0$ the particle density falls off gradually,
\begin{equation}
n_{s}(\mu)\approx\frac{\left(  2m_0\right)  ^{3/2}\Delta_{0}^{2}}{16\pi
\sqrt{|\mu|}}.
\label{ns-tail}
\end{equation}
At $\mu=0$, we find $n_{s}(0)/n(\Delta_{0}) =\frac{1}{2}K\left(
  \frac{1}{2}\right) \approx0.927$.

Unlike the normal case, $n_s(\mu)$ does not vanish at $\mu=0$.  More importantly,
one can see from Eq.~(\ref{eq:G}) that the function $n_s(\mu)$ is analytic at all
$\mu$.  This indicates that the Lifshitz transition at $\mu=0$ is completely smeared
by the superconductivity.  Thus, in the thermodynamic sense, the change between the
behaviors of the system at positive and negative values of the chemical potential
should be classified as a crossover.

\begin{figure}[tb]
\vspace{-0.1in}
\begin{center}
\includegraphics[width=3.0in]{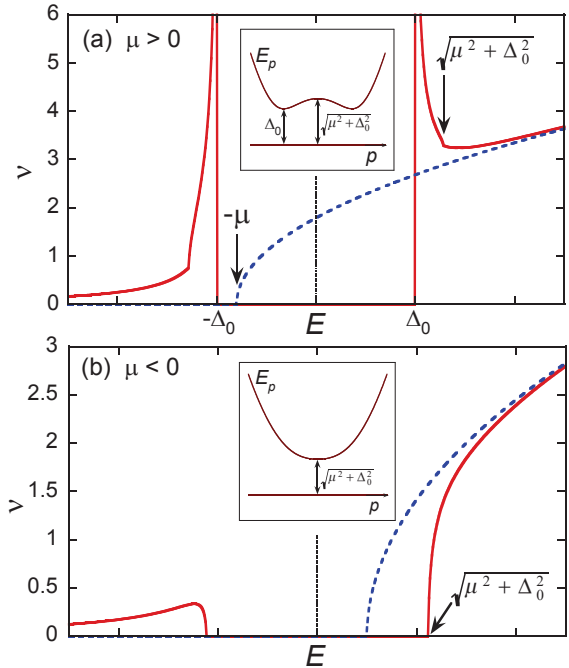}
\vspace{-0.13in}
\end{center}
\caption{Typical shapes of the DoS $\nu_s(E)$, Eq.~(\ref{DoSResult}),
  for the shallow band in superconducting state for (a) positive
  chemical potential, $\mu=0.8\Delta_{0}$, and (b) negative chemical
  potential, $\mu=-0.5\Delta_{0}$.  For comparison, the corresponding
  normal-state DoS $\nu_n(E)$, Eq.~(\ref{eq:normal_DoS}), is also
  shown by the dashed lines. The unit of DoS is $\nu_n(\Delta_0)$. The
  insets illustrate shapes of quasiparticle spectra.}
\label{Fig-DoSshape}
\vspace{-0.18in}
\end{figure}

On the other hand, the spectrum of quasiparticles changes qualitatively at the
normal-state Lifshitz transition point, $\mu=0$, see insets in
Fig.~\ref{Fig-DoSshape}. For $\mu>0$ the gap in the spectrum, $E_{g} =\Delta_{0}$ is
realized at the Fermi momentum $p=p_{F}=\sqrt{2m_0\mu}$, whereas for $\mu<0$ the
spectral gap $E_{g}=\sqrt{\Delta_{0}^{2}+\mu^{2}}$ is at the band center $p=0$. This
has dramatic consequences for the behavior of the density of states of the system.

The shallow band contribution to the DoS is given by
\begin{equation}
\nu_{s}(E)=\int\frac{d^{3}p}{(2\pi)^{3}}\frac{1}{2}\left(  1+\frac{\xi_{p}}
{E}\right)  \delta\left(  |E|-E_{p}\right)  .\label{DoSFormula}
\end{equation}
Here the electron and hole parts of DoS correspond to the energy regions $E>0$ and
$E<0$, respectively. The momentum integral is determined by the roots of the
equation $\left(
  p^{2}/2m_0\!-\!\mu\right) ^{2}\!+\Delta_{0} ^{2}\!=E^{2}$.
The resulting DoS has the form
\begin{align}
\nu_{s}(E)  &  =\frac{\left(  2m_0\right)  ^{3/2}}{8\pi^{2}}\operatorname{Re}
\left[  \sum_{\delta=\pm1}\left(  \frac{|E|}{\sqrt{E^{2}-\Delta_{0}^{2}}
}+\mathrm{sign}(E)\delta\right)  \right. \nonumber\\
&  \times\left.  \sqrt{\mu+\delta\sqrt{E^{2}-\Delta_{0}^{2}}}\right]  .
\label{DoSResult}
\end{align}
Note that the term with $\delta=-1$ contributes to Eq.~(\ref{DoSResult}) only if
$\mu>0$ and $|E|<\sqrt {\mu^{2}+\Delta_{0}^{2}}$.  At $\Delta_0=0$ our result
(\ref{DoSResult}) recovers the normal state DoS
\begin{equation}
  \label{eq:normal_DoS}
  \nu_{n}(E)=\frac{\left(  2m_0\right)  ^{3/2}}{4\pi^{2}}\sqrt{E+\mu}\ \theta(E+\mu).
\end{equation}
Representative DoSs for positive and negative $\mu$ are shown in
Fig.~\ref{Fig-DoSshape}.

Despite its simplicity, the result (\ref{DoSResult}) has a number of interesting
features.  As expected, in the limit $\mu\gg\Delta_{0}$ the DoS approaches the
standard symmetric BCS shape
\[
\nu_{s}(E)\approx\nu_n(0)\frac{|E|}{\sqrt{E^{2}-\Delta_{0}^{2}}} \ \text{for
}\Delta_{0}<|E|\ll\mu.
\]
The above result also describes the main diverging term for
$E\rightarrow\pm\Delta_{0}$, meaning, in particular, that it remains symmetric for
any positive $\mu$. Nevertheless, in the region $\mu\sim\Delta_{0}$, due to the
violation of the particle-hole symmetry, the overall DoS shape acquires significant
asymmetry, see Fig.\ \ref{Fig-DoSshape}(a). In contrast to the normal-state DoS,
which terminates at $E=-\mu$, the superconducting DoS remains finite at negative
energies $E<-E_g$. In particular, at $-E\gg\sqrt{\mu^{2}+\Delta_{0}^{2}}$ it has a
power-law tail
\begin{equation}
\nu_{s}(E)\approx\frac{\left(  2m_0\right)  ^{3/2}}{8\pi^{2}}\frac{\Delta
_{0}^{2}}{2|E|^{3/2}}. \label{Tail}
\end{equation}
Another peculiar feature of the DoS at $\mu>0$ is the square-root singularity at the
energies $E=\pm\sqrt{\mu^{2}+\Delta_{0}^{2}}$ marked in Fig.\ \ref{Fig-DoSshape}(a).
This singularity appears due to the quasiparticle band edge at $p=0$, see inset in
Fig.\ \ref{Fig-DoSshape}(a).

The DoS exhibits a qualitatively different behavior at $\mu<0$.  In this regime it
no longer diverges at the spectral gap energies $E=\pm E_{g}$, but approaches zero
as $\sqrt{|E|-E_{g}}$, similar to behavior at the band edge in the normal state, see
Fig.~\ref{Fig-DoSshape}(b). More precisely, for $E\rightarrow \pm E_{g}$
\[
\nu_{s}(E) \! \approx\!\frac{\left(  2m_0\right)  ^{3/2}}{8\pi^{2}}\left(
\frac{E_{g}}{|\mu|}\pm1\right)
\sqrt{\frac{E_{g}}{|\mu|}\left(  |E|\!-\!E_{g}\right)  }.
\]

The qualitative difference in the behavior of the DoS at positive and negative $\mu$
is a direct consequence of the change in the excitation spectrum shown in the insets
of Figs.~\ref{Fig-DoSshape}(a) and \ref{Fig-DoSshape}(b).  At $\mu>0$ the
square-root singularity of the DoS at $E\to\pm E_g$ is due to the fact that the
lowest energy quasiparticle state has momentum $p\neq0$.  At $\mu<0$ the minimum of
the excitation spectrum is at $p=0$, resulting in vanishing density of states at
$E\to\pm E_g$.  Thus, a careful measurement of the density of states at different
values of the chemical potential should reveal a well-defined ``crossover point''
separating the regimes illustrated in the two panels of Fig.~\ref{Fig-DoSshape}.

Our discussion of the superconductivity in the shallow band neglected the possible
momentum dependence of the pair hopping amplitudes $V_{0,j}$ and the resulting
dependence of the pairing amplitude $\Delta_0(p)$ in the Hamiltonian (\ref{hamilt}).
Such dependence is indeed weak because the characteristic scale of the dependence
$V_{0,j}(p)$ is of the order of the large Fermi momentum in the deep band $j$.  It
is easy to show that a weak dependence of $\Delta_0$ on momentum will result in a
shift of the ``crossover point'' separating the regimes of
Fig.~\ref{Fig-DoSshape}(a) and Fig.~\ref{Fig-DoSshape}(b) away from $\mu=0$
\footnote{If the  $p$ dependence of $\Delta_0$ is taken into account, the minimum
spectral gap shifts to the band center, $p=0$, at $\mu=
m_0\Delta_0(0)\,d^2\!\Delta_0/dp^2$.}.

Recent papers \cite{LubashevskyNat2012,miao_2014,OkazakiSciRep14} studied the
spectrum of excitations in the shallow band of iron-based superconductors and
discovered that the minimum energy is achieved at $p=0$.  This observation is
consistent with the scenario shown in the inset of Fig.~\ref{Fig-DoSshape}(b).  The
authors of Refs.~\cite{LubashevskyNat2012,miao_2014,OkazakiSciRep14} interpreted
this observation as a possible evidence of the Bose-Einstein condensation scenario
of superconductivity. The latter assumes that two electrons in an otherwise empty
shallow band form a bound state. In three dimensions such binding of electrons in
pairs requires strong attractive interaction between them.  In all other
superconductors explored to date, the minimum of the excitation spectrum is achieved
at $p\neq0$, indicating that the interactions are weak, and electron pairing instead
follows the conventional BCS scenario.  Our work shows that the behavior shown in
Fig.~\ref{Fig-DoSshape}(b) may also be observed in multiband BCS superconductors due
to pair hopping into the shallow band.

It was recently reported  that in the compounds LiFeAs \cite{UmezawaPRL12} and
LiFe$_{1-x}$Co$_{x}$As \cite{miao_2014} the shallow band has the larger gap than the
deep bands, which have conventional quasiparticle spectra, $|\Delta_0|>|\Delta_j|$.
We point out that this does not contradict the scenario of induced superconductivity
in the shallow band. For instance, in the case of just one deep band, one can easily
obtain using Eq.~(\ref{ShallGapEq}) that $\Delta_0/\Delta_1=V_{0,1}/V_{1,1}$.  It is
natural to expect that all paring amplitudes are of the same order of magnitude.
Thus there is no reason why a situation with $|V_{0,1}|>V_{1,1}$ may not be
realized, in which case $|\Delta_0|$ would exceed $|\Delta_1|$. Note that even in
this regime the shallow band still gives a negligible contribution to
superconducting pairing because $\nu_0\ll \nu_1$. It is straightforward to
generalize the above argument to the case of several deep bands. On the other hand,
the predicted shapes of the DoS illustrated in Fig.\ \ref{Fig-DoSshape} are most
easily observed in materials where $\Delta_0$ is the smallest gap.  In this case,
the singular behavior at energies near $\pm\Delta_0$ is not obscured by the
nonvanishing contributions to the DoS from the deep bands.

To summarize, we showed that the Lifshitz transition in multiband metals with a
shallow band is smeared by superconductivity.  In particular, the particle density
varies continuously as a function of the chemical potential, as shown in
Fig.~\ref{Fig-n-mu}.  The resulting crossover is nevertheless characterized by
qualitatively different behaviors of the density of states above and below certain
value of $\mu$, as illustrated in the two panels of Fig.~\ref{Fig-DoSshape}.

The authors are grateful to A. Kanigel and A. A. Varlamov for discussions.  This
work was supported by the U.S. Department of Energy, Office of Science, Materials
Sciences and Engineering Division.


\bibliography{LifshitzMultiSupercond}

\begin{thebibliography}{22}%
\makeatletter
\providecommand \@ifxundefined [1]{%
 \@ifx{#1\undefined}
}%
\providecommand \@ifnum [1]{%
 \ifnum #1\expandafter \@firstoftwo
 \else \expandafter \@secondoftwo
 \fi
}%
\providecommand \@ifx [1]{%
 \ifx #1\expandafter \@firstoftwo
 \else \expandafter \@secondoftwo
 \fi
}%
\providecommand \natexlab [1]{#1}%
\providecommand \enquote  [1]{``#1''}%
\providecommand \bibnamefont  [1]{#1}%
\providecommand \bibfnamefont [1]{#1}%
\providecommand \citenamefont [1]{#1}%
\providecommand \href@noop [0]{\@secondoftwo}%
\providecommand \href [0]{\begingroup \@sanitize@url \@href}%
\providecommand \@href[1]{\@@startlink{#1}\@@href}%
\providecommand \@@href[1]{\endgroup#1\@@endlink}%
\providecommand \@sanitize@url [0]{\catcode `\\12\catcode `\$12\catcode
  `\&12\catcode `\#12\catcode `\^12\catcode `\_12\catcode `\%12\relax}%
\providecommand \@@startlink[1]{}%
\providecommand \@@endlink[0]{}%
\providecommand \url  [0]{\begingroup\@sanitize@url \@url }%
\providecommand \@url [1]{\endgroup\@href {#1}{\urlprefix }}%
\providecommand \urlprefix  [0]{URL }%
\providecommand \Eprint [0]{\href }%
\providecommand \doibase [0]{http://dx.doi.org/}%
\providecommand \selectlanguage [0]{\@gobble}%
\providecommand \bibinfo  [0]{\@secondoftwo}%
\providecommand \bibfield  [0]{\@secondoftwo}%
\providecommand \translation [1]{[#1]}%
\providecommand \BibitemOpen [0]{}%
\providecommand \bibitemStop [0]{}%
\providecommand \bibitemNoStop [0]{.\EOS\space}%
\providecommand \EOS [0]{\spacefactor3000\relax}%
\providecommand \BibitemShut  [1]{\csname bibitem#1\endcsname}%
\let\auto@bib@innerbib\@empty
\bibitem [{\citenamefont {Kamihara}\ \emph {et~al.}(2008)\citenamefont
  {Kamihara}, \citenamefont {Watanabe}, \citenamefont {Hirano},\ and\
  \citenamefont {Hosono}}]{KamiharaJACS08}%
  \BibitemOpen
  \bibfield  {author} {\bibinfo {author} {\bibfnamefont {Y.}~\bibnamefont
  {Kamihara}}, \bibinfo {author} {\bibfnamefont {T.}~\bibnamefont {Watanabe}},
  \bibinfo {author} {\bibfnamefont {M.}~\bibnamefont {Hirano}}, \ and\ \bibinfo
  {author} {\bibfnamefont {H.}~\bibnamefont {Hosono}},\ }\href {\doibase
  10.1021/ja800073m} {\bibfield  {journal} {\bibinfo  {journal} {Journ. Amer.
  Chem. Soc.}\ }\textbf {\bibinfo {volume} {130}},\ \bibinfo {pages} {3296}
  (\bibinfo {year} {2008})}\BibitemShut {NoStop}%
\bibitem [{\citenamefont {Paglione}\ and\ \citenamefont
  {Greene}(2010)}]{PaglioneNPhys10}%
  \BibitemOpen
  \bibfield  {author} {\bibinfo {author} {\bibfnamefont {J.}~\bibnamefont
  {Paglione}}\ and\ \bibinfo {author} {\bibfnamefont {R.~L.}\ \bibnamefont
  {Greene}},\ }\href {\doibase 10.1038/nphys1759} {\bibfield  {journal}
  {\bibinfo  {journal} {Nat Phys}\ }\textbf {\bibinfo {volume} {6}},\ \bibinfo
  {pages} {645} (\bibinfo {year} {2010})}\BibitemShut {NoStop}%
\bibitem [{\citenamefont {Johnston}(2010)}]{JohnstonAdvPhys10}%
  \BibitemOpen
  \bibfield  {author} {\bibinfo {author} {\bibfnamefont {D.~C.}\ \bibnamefont
  {Johnston}},\ }\href {\doibase 10.1080/00018732.2010.513480} {\bibfield
  {journal} {\bibinfo  {journal} {Adv. Phys.}\ }\textbf {\bibinfo {volume}
  {59}},\ \bibinfo {pages} {803} (\bibinfo {year} {2010})}\BibitemShut
  {NoStop}%
\bibitem [{\citenamefont {Hirschfeld}\ \emph {et~al.}(2011)\citenamefont
  {Hirschfeld}, \citenamefont {Korshunov},\ and\ \citenamefont
  {Mazin}}]{HirschfeldRPP11}%
  \BibitemOpen
  \bibfield  {author} {\bibinfo {author} {\bibfnamefont {P.~J.}\ \bibnamefont
  {Hirschfeld}}, \bibinfo {author} {\bibfnamefont {M.~M.}\ \bibnamefont
  {Korshunov}}, \ and\ \bibinfo {author} {\bibfnamefont {I.~I.}\ \bibnamefont
  {Mazin}},\ }\href {http://stacks.iop.org/0034-4885/74/i=12/a=124508}
  {\bibfield  {journal} {\bibinfo  {journal} {Rep. Progr. Phys.}\ }\textbf
  {\bibinfo {volume} {74}},\ \bibinfo {pages} {124508} (\bibinfo {year}
  {2011})}\BibitemShut {NoStop}%
\bibitem [{\citenamefont {Chubukov}(2012)}]{ChubukovAnnual12}%
  \BibitemOpen
  \bibfield  {author} {\bibinfo {author} {\bibfnamefont {A.}~\bibnamefont
  {Chubukov}},\ }\href {\doibase 10.1146/annurev-conmatphys-020911-125055}
  {\bibfield  {journal} {\bibinfo  {journal} {Ann. Rev. of Cond. Mat. Phys.}\
  }\textbf {\bibinfo {volume} {3}},\ \bibinfo {pages} {57} (\bibinfo {year}
  {2012})}\BibitemShut {NoStop}%
\bibitem [{\citenamefont {Liu}\ \emph {et~al.}(2010)\citenamefont {Liu},
  \citenamefont {Kondo}, \citenamefont {Fernandes}, \citenamefont {Palczewski},
  \citenamefont {Mun}, \citenamefont {Ni}, \citenamefont {Thaler},
  \citenamefont {Bostwick}, \citenamefont {Rotenberg}, \citenamefont
  {Schmalian}, \citenamefont {Bud’ko}, \citenamefont {Canfield},\ and\
  \citenamefont {Kaminski}}]{LiuNatPhys10}%
  \BibitemOpen
  \bibfield  {author} {\bibinfo {author} {\bibfnamefont {C.}~\bibnamefont
  {Liu}}, \bibinfo {author} {\bibfnamefont {T.}~\bibnamefont {Kondo}}, \bibinfo
  {author} {\bibfnamefont {R.~M.}\ \bibnamefont {Fernandes}}, \bibinfo {author}
  {\bibfnamefont {A.~D.}\ \bibnamefont {Palczewski}}, \bibinfo {author}
  {\bibfnamefont {E.~D.}\ \bibnamefont {Mun}}, \bibinfo {author} {\bibfnamefont
  {N.}~\bibnamefont {Ni}}, \bibinfo {author} {\bibfnamefont {A.~N.}\
  \bibnamefont {Thaler}}, \bibinfo {author} {\bibfnamefont {A.}~\bibnamefont
  {Bostwick}}, \bibinfo {author} {\bibfnamefont {E.}~\bibnamefont {Rotenberg}},
  \bibinfo {author} {\bibfnamefont {J.}~\bibnamefont {Schmalian}}, \bibinfo
  {author} {\bibfnamefont {S.~L.}\ \bibnamefont {Bud’ko}}, \bibinfo {author}
  {\bibfnamefont {P.~C.}\ \bibnamefont {Canfield}}, \ and\ \bibinfo {author}
  {\bibfnamefont {A.}~\bibnamefont {Kaminski}},\ }\href {\doibase
  10.1038/nphys1656} {\bibfield  {journal} {\bibinfo  {journal} {Nat. Phys.}\
  }\textbf {\bibinfo {volume} {6}},\ \bibinfo {pages} {419} (\bibinfo {year}
  {2010})}\BibitemShut {NoStop}%
\bibitem [{\citenamefont {Liu}\ \emph {et~al.}(2011)\citenamefont {Liu},
  \citenamefont {Palczewski}, \citenamefont {Dhaka}, \citenamefont {Kondo},
  \citenamefont {Fernandes}, \citenamefont {Mun}, \citenamefont {Hodovanets},
  \citenamefont {Thaler}, \citenamefont {Schmalian}, \citenamefont {Bud'ko},
  \citenamefont {Canfield},\ and\ \citenamefont {Kaminski}}]{LiuPhysRevB11}%
  \BibitemOpen
  \bibfield  {author} {\bibinfo {author} {\bibfnamefont {C.}~\bibnamefont
  {Liu}}, \bibinfo {author} {\bibfnamefont {A.~D.}\ \bibnamefont {Palczewski}},
  \bibinfo {author} {\bibfnamefont {R.~S.}\ \bibnamefont {Dhaka}}, \bibinfo
  {author} {\bibfnamefont {T.}~\bibnamefont {Kondo}}, \bibinfo {author}
  {\bibfnamefont {R.~M.}\ \bibnamefont {Fernandes}}, \bibinfo {author}
  {\bibfnamefont {E.~D.}\ \bibnamefont {Mun}}, \bibinfo {author} {\bibfnamefont
  {H.}~\bibnamefont {Hodovanets}}, \bibinfo {author} {\bibfnamefont {A.~N.}\
  \bibnamefont {Thaler}}, \bibinfo {author} {\bibfnamefont {J.}~\bibnamefont
  {Schmalian}}, \bibinfo {author} {\bibfnamefont {S.~L.}\ \bibnamefont
  {Bud'ko}}, \bibinfo {author} {\bibfnamefont {P.~C.}\ \bibnamefont
  {Canfield}}, \ and\ \bibinfo {author} {\bibfnamefont {A.}~\bibnamefont
  {Kaminski}},\ }\href {\doibase 10.1103/PhysRevB.84.020509} {\bibfield
  {journal} {\bibinfo  {journal} {Phys. Rev. B}\ }\textbf {\bibinfo {volume}
  {84}},\ \bibinfo {pages} {020509} (\bibinfo {year} {2011})}\BibitemShut
  {NoStop}%
\bibitem [{\citenamefont {Miao}\ \emph {et~al.}(2014)\citenamefont {Miao},
  \citenamefont {Qian}, \citenamefont {Shi}, \citenamefont {Richard},
  \citenamefont {Kim}, \citenamefont {Hoesch}, \citenamefont {Xing},
  \citenamefont {Wang}, \citenamefont {Jin}, \citenamefont {Hu},\ and\
  \citenamefont {Ding}}]{miao_2014}%
  \BibitemOpen
  \bibfield  {author} {\bibinfo {author} {\bibfnamefont {H.}~\bibnamefont
  {Miao}}, \bibinfo {author} {\bibfnamefont {T.}~\bibnamefont {Qian}}, \bibinfo
  {author} {\bibfnamefont {X.}~\bibnamefont {Shi}}, \bibinfo {author}
  {\bibfnamefont {P.}~\bibnamefont {Richard}}, \bibinfo {author} {\bibfnamefont
  {T.~K.}\ \bibnamefont {Kim}}, \bibinfo {author} {\bibfnamefont
  {M.}~\bibnamefont {Hoesch}}, \bibinfo {author} {\bibfnamefont {L.~Y.}\
  \bibnamefont {Xing}}, \bibinfo {author} {\bibfnamefont {X.~C.}\ \bibnamefont
  {Wang}}, \bibinfo {author} {\bibfnamefont {C.~Q.}\ \bibnamefont {Jin}},
  \bibinfo {author} {\bibfnamefont {J.~P.}\ \bibnamefont {Hu}}, \ and\ \bibinfo
  {author} {\bibfnamefont {H.}~\bibnamefont {Ding}},\ }\href
  {http://arxiv.org/abs/1406.0983} {\bibfield  {journal} {\bibinfo  {journal}
  {arXiv:1406.0983}\ } (\bibinfo {year} {2014})}\BibitemShut {NoStop}%
\bibitem [{\citenamefont {Lifshitz}(1960)}]{Lifshitz}%
  \BibitemOpen
  \bibfield  {author} {\bibinfo {author} {\bibfnamefont {I.~M.}\ \bibnamefont
  {Lifshitz}},\ }\href@noop {} {\bibfield  {journal} {\bibinfo  {journal} {Zh.
  Eksp. Teor. Fiz.}\ }\textbf {\bibinfo {volume} {38}},\ \bibinfo {pages}
  {1569} (\bibinfo {year} {1960})},\ \bibinfo {note} {[Sov. Phys. JETP
  \textbf{11},1130 (1960)]}\BibitemShut {NoStop}%
\bibitem [{\citenamefont {Blanter}\ \emph {et~al.}(1994)\citenamefont
  {Blanter}, \citenamefont {Kaganov}, \citenamefont {Pantsulaya},\ and\
  \citenamefont {Varlamov}}]{BlanterPhysRep94}%
  \BibitemOpen
  \bibfield  {author} {\bibinfo {author} {\bibfnamefont {Y.}~\bibnamefont
  {Blanter}}, \bibinfo {author} {\bibfnamefont {M.}~\bibnamefont {Kaganov}},
  \bibinfo {author} {\bibfnamefont {A.}~\bibnamefont {Pantsulaya}}, \ and\
  \bibinfo {author} {\bibfnamefont {A.}~\bibnamefont {Varlamov}},\ }\href
  {http://www.sciencedirect.com/science/article/pii/0370157394901031}
  {\bibfield  {journal} {\bibinfo  {journal} {Phys. Rep.}\ }\textbf {\bibinfo
  {volume} {245}},\ \bibinfo {pages} {159 } (\bibinfo {year}
  {1994})}\BibitemShut {NoStop}%
\bibitem [{\citenamefont {Lubashevsky}\ \emph {et~al.}(2012)\citenamefont
  {Lubashevsky}, \citenamefont {Lahoud}, \citenamefont {Chashka}, \citenamefont
  {Podolsky},\ and\ \citenamefont {Kanigel}}]{LubashevskyNat2012}%
  \BibitemOpen
  \bibfield  {author} {\bibinfo {author} {\bibfnamefont {Y.}~\bibnamefont
  {Lubashevsky}}, \bibinfo {author} {\bibfnamefont {E.}~\bibnamefont {Lahoud}},
  \bibinfo {author} {\bibfnamefont {K.}~\bibnamefont {Chashka}}, \bibinfo
  {author} {\bibfnamefont {D.}~\bibnamefont {Podolsky}}, \ and\ \bibinfo
  {author} {\bibfnamefont {A.}~\bibnamefont {Kanigel}},\ }\href {\doibase
  10.1038/nphys2216} {\bibfield  {journal} {\bibinfo  {journal} {Nat. Phys.}\
  }\textbf {\bibinfo {volume} {8}},\ \bibinfo {pages} {309} (\bibinfo {year}
  {2012})}\BibitemShut {NoStop}%
\bibitem [{\citenamefont {Okazaki}\ \emph {et~al.}(2014)\citenamefont
  {Okazaki}, \citenamefont {Ito}, \citenamefont {Ota}, \citenamefont {Kotani},
  \citenamefont {Shimojima}, \citenamefont {Kiss}, \citenamefont {Watanabe},
  \citenamefont {Chen}, \citenamefont {Niitaka}, \citenamefont {Hanaguri},
  \citenamefont {Takagi}, \citenamefont {Chainani},\ and\ \citenamefont
  {Shin}}]{OkazakiSciRep14}%
  \BibitemOpen
  \bibfield  {author} {\bibinfo {author} {\bibfnamefont {K.}~\bibnamefont
  {Okazaki}}, \bibinfo {author} {\bibfnamefont {Y.}~\bibnamefont {Ito}},
  \bibinfo {author} {\bibfnamefont {Y.}~\bibnamefont {Ota}}, \bibinfo {author}
  {\bibfnamefont {Y.}~\bibnamefont {Kotani}}, \bibinfo {author} {\bibfnamefont
  {T.}~\bibnamefont {Shimojima}}, \bibinfo {author} {\bibfnamefont
  {T.}~\bibnamefont {Kiss}}, \bibinfo {author} {\bibfnamefont {S.}~\bibnamefont
  {Watanabe}}, \bibinfo {author} {\bibfnamefont {C.-T.}\ \bibnamefont {Chen}},
  \bibinfo {author} {\bibfnamefont {S.}~\bibnamefont {Niitaka}}, \bibinfo
  {author} {\bibfnamefont {T.}~\bibnamefont {Hanaguri}}, \bibinfo {author}
  {\bibfnamefont {H.}~\bibnamefont {Takagi}}, \bibinfo {author} {\bibfnamefont
  {A.}~\bibnamefont {Chainani}}, \ and\ \bibinfo {author} {\bibfnamefont
  {S.}~\bibnamefont {Shin}},\ }\href
  {http://www.nature.com/srep/2014/140227/srep04109/full/srep04109.html}
  {\bibfield  {journal} {\bibinfo  {journal} {Sci. Rep.}\ }\textbf {\bibinfo
  {volume} {4}},\ \bibinfo {pages} {1} (\bibinfo {year} {2014})}\BibitemShut
  {NoStop}%
\bibitem [{\citenamefont {Giorgini}\ \emph {et~al.}(2008)\citenamefont
  {Giorgini}, \citenamefont {Pitaevskii},\ and\ \citenamefont
  {Stringari}}]{GiorginiRevModPhys08}%
  \BibitemOpen
  \bibfield  {author} {\bibinfo {author} {\bibfnamefont {S.}~\bibnamefont
  {Giorgini}}, \bibinfo {author} {\bibfnamefont {L.~P.}\ \bibnamefont
  {Pitaevskii}}, \ and\ \bibinfo {author} {\bibfnamefont {S.}~\bibnamefont
  {Stringari}},\ }\href {\doibase 10.1103/RevModPhys.80.1215} {\bibfield
  {journal} {\bibinfo  {journal} {Rev. Mod. Phys.}\ }\textbf {\bibinfo {volume}
  {80}},\ \bibinfo {pages} {1215} (\bibinfo {year} {2008})}\BibitemShut
  {NoStop}%
\bibitem [{\citenamefont {Makarov}\ and\ \citenamefont
  {Bar'yakhtar}(1965)}]{MakarovBaryakhtarZheTF65}%
  \BibitemOpen
  \bibfield  {author} {\bibinfo {author} {\bibfnamefont {V.~I.}\ \bibnamefont
  {Makarov}}\ and\ \bibinfo {author} {\bibfnamefont {V.~G.}\ \bibnamefont
  {Bar'yakhtar}},\ }\href@noop {} {\bibfield  {journal} {\bibinfo  {journal}
  {Zh. Eksp. Teor. Fiz.}\ }\textbf {\bibinfo {volume} {48}},\ \bibinfo {pages}
  {1717} (\bibinfo {year} {1965})},\ \bibinfo {note} {[Sov. Phys. JETP
  \textbf{21},1151 (1965)]}\BibitemShut {NoStop}%
\bibitem [{\citenamefont {Suhl}\ \emph {et~al.}(1959)\citenamefont {Suhl},
  \citenamefont {Matthias},\ and\ \citenamefont {Walker}}]{MultibandBCS}%
  \BibitemOpen
  \bibfield  {author} {\bibinfo {author} {\bibfnamefont {H.}~\bibnamefont
  {Suhl}}, \bibinfo {author} {\bibfnamefont {B.~T.}\ \bibnamefont {Matthias}},
  \ and\ \bibinfo {author} {\bibfnamefont {L.~R.}\ \bibnamefont {Walker}},\
  }\href@noop {} {\bibfield  {journal} {\bibinfo  {journal} {Phys. Rev. Lett.}\
  }\textbf {\bibinfo {volume} {3}},\ \bibinfo {pages} {552} (\bibinfo {year}
  {1959})}\BibitemShut {NoStop}%
\bibitem [{\citenamefont {Moskalenko}(1959)}]{MoskalenkoFMM59}%
  \BibitemOpen
  \bibfield  {author} {\bibinfo {author} {\bibfnamefont {V.~A.}\ \bibnamefont
  {Moskalenko}},\ }\href@noop {} {\bibfield  {journal} {\bibinfo  {journal}
  {Fiz. Met. Metalloved.}\ }\textbf {\bibinfo {volume} {8}},\ \bibinfo {pages}
  {503} (\bibinfo {year} {1959})}\BibitemShut {NoStop}%
\bibitem [{\citenamefont {Geilikman}\ \emph {et~al.}(1967)\citenamefont
  {Geilikman}, \citenamefont {Zaitsev},\ and\ \citenamefont
  {Kresin}}]{GeilikmanFTT67}%
  \BibitemOpen
  \bibfield  {author} {\bibinfo {author} {\bibfnamefont {B.}~\bibnamefont
  {Geilikman}}, \bibinfo {author} {\bibfnamefont {R.}~\bibnamefont {Zaitsev}},
  \ and\ \bibinfo {author} {\bibfnamefont {V.}~\bibnamefont {Kresin}},\
  }\href@noop {} {\bibfield  {journal} {\bibinfo  {journal} {Fiz. Tverd. Tela}\
  }\textbf {\bibinfo {volume} {9}},\ \bibinfo {pages} {821} (\bibinfo {year}
  {1967})},\ \bibinfo {note} {[Sov. Phys. Solid State \textbf{9}, 642
  (1967)]}\BibitemShut {NoStop}%
\bibitem [{Note1()}]{Note1}%
  \BibitemOpen
  \bibinfo {note} {The scenario we consider here is distinctly different from
  the situation when the shallow band dominates the Cooper pairing so that its
  depletion destroys superconducting state \cite
  {LiuPhysRevB11,CooleyPhysRevLett05,InnocentiPhysRevB10}.}\BibitemShut {Stop}%
\bibitem [{Note2()}]{Note2}%
  \BibitemOpen
  \bibinfo {note} {If the $p$ dependence of $\Delta _0$ is taken into account,
  the minimum spectral gap shifts to the band center, $p=0$, at $\mu =
  m_0\Delta _0(0)\protect \tmspace +\thinmuskip {.1667em}d^2\protect \tmspace
  -\thinmuskip {.1667em}\Delta _0/dp^2$.}\BibitemShut {Stop}%
\bibitem [{\citenamefont {Umezawa}\ \emph {et~al.}(2012)\citenamefont
  {Umezawa}, \citenamefont {Li}, \citenamefont {Miao}, \citenamefont
  {Nakayama}, \citenamefont {Liu}, \citenamefont {Richard}, \citenamefont
  {Sato}, \citenamefont {He}, \citenamefont {Wang}, \citenamefont {Chen},
  \citenamefont {Ding}, \citenamefont {Takahashi},\ and\ \citenamefont
  {Wang}}]{UmezawaPRL12}%
  \BibitemOpen
  \bibfield  {author} {\bibinfo {author} {\bibfnamefont {K.}~\bibnamefont
  {Umezawa}}, \bibinfo {author} {\bibfnamefont {Y.}~\bibnamefont {Li}},
  \bibinfo {author} {\bibfnamefont {H.}~\bibnamefont {Miao}}, \bibinfo {author}
  {\bibfnamefont {K.}~\bibnamefont {Nakayama}}, \bibinfo {author}
  {\bibfnamefont {Z.-H.}\ \bibnamefont {Liu}}, \bibinfo {author} {\bibfnamefont
  {P.}~\bibnamefont {Richard}}, \bibinfo {author} {\bibfnamefont
  {T.}~\bibnamefont {Sato}}, \bibinfo {author} {\bibfnamefont {J.~B.}\
  \bibnamefont {He}}, \bibinfo {author} {\bibfnamefont {D.-M.}\ \bibnamefont
  {Wang}}, \bibinfo {author} {\bibfnamefont {G.~F.}\ \bibnamefont {Chen}},
  \bibinfo {author} {\bibfnamefont {H.}~\bibnamefont {Ding}}, \bibinfo {author}
  {\bibfnamefont {T.}~\bibnamefont {Takahashi}}, \ and\ \bibinfo {author}
  {\bibfnamefont {S.-C.}\ \bibnamefont {Wang}},\ }\href {\doibase
  10.1103/PhysRevLett.108.037002} {\bibfield  {journal} {\bibinfo  {journal}
  {Phys. Rev. Lett.}\ }\textbf {\bibinfo {volume} {108}},\ \bibinfo {pages}
  {037002} (\bibinfo {year} {2012})}\BibitemShut {NoStop}%
\bibitem [{\citenamefont {Cooley}\ \emph {et~al.}(2005)\citenamefont {Cooley},
  \citenamefont {Zambano}, \citenamefont {Moodenbaugh}, \citenamefont {Klie},
  \citenamefont {Zheng},\ and\ \citenamefont {Zhu}}]{CooleyPhysRevLett05}%
  \BibitemOpen
  \bibfield  {author} {\bibinfo {author} {\bibfnamefont {L.~D.}\ \bibnamefont
  {Cooley}}, \bibinfo {author} {\bibfnamefont {A.~J.}\ \bibnamefont {Zambano}},
  \bibinfo {author} {\bibfnamefont {A.~R.}\ \bibnamefont {Moodenbaugh}},
  \bibinfo {author} {\bibfnamefont {R.~F.}\ \bibnamefont {Klie}}, \bibinfo
  {author} {\bibfnamefont {J.-C.}\ \bibnamefont {Zheng}}, \ and\ \bibinfo
  {author} {\bibfnamefont {Y.}~\bibnamefont {Zhu}},\ }\href {\doibase
  10.1103/PhysRevLett.95.267002} {\bibfield  {journal} {\bibinfo  {journal}
  {Phys. Rev. Lett.}\ }\textbf {\bibinfo {volume} {95}},\ \bibinfo {pages}
  {267002} (\bibinfo {year} {2005})}\BibitemShut {NoStop}%
\bibitem [{\citenamefont {Innocenti}\ \emph {et~al.}(2010)\citenamefont
  {Innocenti}, \citenamefont {Poccia}, \citenamefont {Ricci}, \citenamefont
  {Valletta}, \citenamefont {Caprara}, \citenamefont {Perali},\ and\
  \citenamefont {Bianconi}}]{InnocentiPhysRevB10}%
  \BibitemOpen
  \bibfield  {author} {\bibinfo {author} {\bibfnamefont {D.}~\bibnamefont
  {Innocenti}}, \bibinfo {author} {\bibfnamefont {N.}~\bibnamefont {Poccia}},
  \bibinfo {author} {\bibfnamefont {A.}~\bibnamefont {Ricci}}, \bibinfo
  {author} {\bibfnamefont {A.}~\bibnamefont {Valletta}}, \bibinfo {author}
  {\bibfnamefont {S.}~\bibnamefont {Caprara}}, \bibinfo {author} {\bibfnamefont
  {A.}~\bibnamefont {Perali}}, \ and\ \bibinfo {author} {\bibfnamefont
  {A.}~\bibnamefont {Bianconi}},\ }\href {\doibase 10.1103/PhysRevB.82.184528}
  {\bibfield  {journal} {\bibinfo  {journal} {Phys. Rev. B}\ }\textbf {\bibinfo
  {volume} {82}},\ \bibinfo {pages} {184528} (\bibinfo {year}
  {2010})}\BibitemShut {NoStop}%
\end{thebibliography}%

\end{document}